# Prediction of the Atomization Energy of Molecules Using Coulomb Matrix and Atomic Composition in a Bayesian Regularized Neural Networks

Alain B. Tchagang and Julio J. Valdés

National Research Council Canada
Digital Technologies Research Centre
M-50, 1200 Montréal Road Ottawa ON K4A 0S2, Canada
{alain.tchagang, julio.valdes}@nrc-crnc.gc.ca

**Abstract.** Exact calculation of electronic properties of molecules is a fundamental step for intelligent and rational compounds and materials design. The intrinsically graph-like and non-vectorial nature of molecular data generates a unique and challenging machine learning problem. In this paper we embrace a learning from scratch approach where the quantum mechanical electronic properties of molecules are predicted directly from the raw molecular geometry, similar to some recent works. But, unlike these previous endeavors, our study suggests a benefit from combining molecular geometry embedded in the Coulomb matrix with the atomic composition of molecules. Using the new combined features in a Bayesian regularized neural networks, our results improve well-known results from the literature on the QM7 dataset from a mean absolute error of 3.51 kcal/mol down to 3.0 kcal/mol.

**Keywords:** Atomization Energy, Atomic Composition, Bayesian Regularization, Coulomb Matrix, Electronic Properties, Molecules, Neural Networks.

## 1 Introduction

Finding new molecules, compounds or materials with desired properties is strategic to the innovation and progress of many chemical, agrochemical and pharmaceutical industries. One of the major challenges consists of making quantitative estimates in the chemical compound space at moderate computational cost (milliseconds per compound or faster). Currently only high level quantum-chemistry calculations, which can take days per molecule depending on property and system, yield the desired chemical accuracy of 1 kcal/mol required for computational molecular and material design [1].

Recent technological advances have shown that data-to-knowledge approaches are beginning to show enormous promise within materials science. Intelligent exploration and exploitation of the vast materials property space has the potential to alleviate the cost, risks, and time involved in trial-by-error approach experiment cycles used by current techniques to identify useful compounds [2]. For example, obtaining atomization energies from the Schrödinger equation solver is computationally expensive and, as a consequence, only a fraction of the molecules in the chemical compound space can be labeled. By training a machine learning algorithm on the few label ones, the trained quantum mechanics machine learning (QM/ML) model can be used to generalize from these few data points to unseen molecules [3]. One of the central questions in QM/ML is how to represent molecules in a way that makes prediction of molecular properties feasible and accurate [4]. This question has already been extensively discussed in the cheminformatics literature, and many so-called molecular descriptors exist [5]. Unfortunately, they often require a substantial amount of domain knowledge and engineering. Furthermore, they are not necessarily transferable across the whole chemical compound space [1].

In this paper, we follow a more direct approach introduced in [6], and adopted by several other authors. We learn the mapping between the molecule and its atomization energy from scratch using the Coulomb matrix as a low-level molecular descriptor [6-7]. Coulomb matrix is invariant to translation and rotation but not to permutations or re-indexing of the atoms. Methods to tackle this issue have been proposed. Examples include Coulomb sorted eigen spectrum [1], Coulomb sorted L2 norm of the matrix column [7], Coulomb bag of bonds [8], and random Coulomb matrices [1, 3]. Our study extends the work of [1, 3, 6, 7]. Unlike these previous authors, we show that by combining the molecular geometry embedded in the Coulomb matrix with atomicity or atomic composition of molecules (i.e. atom counts of each type in a molecule), the outcome of the QM/ML models can be significantly improved.

To test this new hypothesis, three representations are constructed: (1) Coulomb matrix, (2) atomic composition of molecules, and (3) the combination of the Coulomb matrix and the atomic composition of molecules, and each one is used as input to a well-defined multilayer Bayesian regularized neural networks [9-13]. Results obtained using the combination of the Coulomb matrix and the atomic composition showed better predictions by a difference of more than 1.5 kcal/mol compared to using either the Coulomb matrix or the atomic composition solely. More interestingly, the mean absolute error (MAE) = 3.0 kcal/mol obtained in this study is lower than the 3.51 kcal/mol well-known results obtained in [1, 3]. These results confirm the efficacy of using the atomic composition of molecules in a QM/ML model for their electronic properties predictions. Furthermore, the Bayesian regularized neural network is shown to be a suitable candidate for the modeling of molecular data.

The rest of this paper is organized as follows. In Section II, the dataset used in this study is described. Section III provides a detailed description of the proposed method. Section IV presents the results and Section V the conclusions.

## 2      Materials

The QM7 dataset used in this study is a subset of the GDB-13 dataset [14]. The version used here is the one published in [7] consisting of 7102 small organic molecules and their associated atomization energy. These molecules are composed of a maximum of 23 atoms. Molecules are converted to a suitable Cartesian coordinates representation using universal forcefield method [15] as implemented in the software OpenBabel [16]. Atomization energies are calculated for each molecule and ranging from −800 to −2000 kcal/mol. Note that all the 7102 molecules are unique and there are no isomers in the set.

## 3   Methods

Coulomb matrix and atomic composition of each molecule are computed using the atomic coordinates and the chemical formulae of each molecule respectively as described in the QM7 dataset. Next, atomic composition and Coulomb matrix are either combined or used separately as input to a regularized Bayesian neural network for the prediction of the atomization energy.

### 3.1   Atomicity, Atom Counts or Atomic Composition

Let's define $\Omega = \{\Omega_1, \Omega_2, ..., \Omega_m, ..., \Omega_M\}$, the set of possible molecules in the chemical compound space (CCS). By construction, this space is very large. In this study, we will assume that it is bounded by M. Let's define A the set of unique atoms that make $\Omega$. A is bounded by K and it is defined as: $A = \{A^1, A^2, ..., A^k, ..., A^K\}$. Let's define a chemical operator "." that combines atoms among them in a specific numbers $\alpha_k^m$ and according to the laws of chemistry to form a stable molecule $\Omega_m$. The chemical formulae of $\Omega_m$ can be written as: $\Omega_m = \alpha_1^m A^1 . \alpha_2^m A^2 .... \alpha_k^m A^k .... \alpha_K^m A^K$, or as in chemical textbook.

$$\Omega_m \equiv A^1_{\alpha_1^m} A^2_{\alpha_2^m} ... A^k_{\alpha_k^m} ... A^K_{\alpha_K^m} \tag{4}$$

The atomic composition (AC) of molecule $\Omega_m$ in the atomic space $[A^1\ A^2\ ...\ A^k\ ...\ A^K]$ is defined as $[\alpha_1^m\ \alpha_2^m\ ...\ \alpha_k^m\ ...\ \alpha_K^m]$, where $\alpha_k^m$ is a positive integer that represents the number of atom $A^k$ in molecule $\Omega_m$. The AC of the M molecules in the atomic space $[A^1\ A^2\ ...\ A^k\ ...\ A^K]$ can be viewed as an M×K matrix α, **Equation 5**.

$$\alpha = \begin{bmatrix} \Omega_1 \\ \Omega_2 \\ \vdots \\ \Omega_m \\ \vdots \\ \Omega_M \end{bmatrix} = \begin{bmatrix} \alpha_1^1 & \alpha_2^1 & ... & \alpha_k^1 & ... & \alpha_K^1 \\ \alpha_1^2 & \alpha_2^2 & ... & \alpha_k^2 & ... & \alpha_K^2 \\ \vdots & \vdots & ... & \vdots & ... & \vdots \\ \alpha_1^m & \alpha_2^m & ... & \alpha_k^m & ... & \alpha_K^m \\ \vdots & ... & ... & \vdots & ... & \vdots \\ \alpha_1^M & \alpha_2^M & ... & \alpha_k^M & ... & \alpha_K^M \end{bmatrix} \tag{5}$$

Row α(m,:) of α corresponds to the AC of the m[th] molecule ($\Omega_m$). Column α(:,k) corresponds to the number of atom $A^k$ in each molecule of $\Omega$. K is an integer and correspond to the number of unique atoms that makes $\Omega$. For example, given a set of seven molecules: $\Omega = \{CH_4, C_2H_2, C_3H_6, C_2NH_3, OC_2H_2, ONC_3H_3, SC_3NH_3\}$. The set of unique atoms that makes $\Omega$ is A = {C, H, N, O, S}. The matrix α is then:

$$\alpha = \begin{bmatrix} CH_4 \\ C_2H_2 \\ C_3H_6 \\ C_2NH_3 \\ OC_2H_2 \\ ONC_3H_3 \\ SC_3NH_3 \end{bmatrix} = \begin{bmatrix} 1 & 4 & 0 & 0 & 0 \\ 2 & 2 & 0 & 0 & 0 \\ 3 & 6 & 0 & 0 & 0 \\ 2 & 3 & 1 & 0 & 0 \\ 2 & 2 & 0 & 1 & 0 \\ 3 & 3 & 1 & 1 & 0 \\ 3 & 3 & 1 & 0 & 1 \end{bmatrix} \qquad (6)$$

It is obvious that this representation is not unique. That is two molecules with identical atomic composition may have different electronic properties. Isomers are great examples in this case. They are compound with the same molecular formulas but that are structurally different in some way, and they can have different chemical, physical and biological properties [17]. It is also worth to note that such molecular representation had been explored in the past in quantitative structure activity relationship and correspond to a different form of the Atomistic index developed by Burden [13].

### 3.2 Coulomb Matrix

The Coulomb matrix (CM) has recently been widely used as molecular descriptors in the QM/ML models [1, 3, 6, 7]. Given a molecule its Coulomb matrix CM = $[c_{ij}]$ is defined by **Equation 1**.

$$c_{ij} = \begin{cases} 0.5 Z_i^{2.4} & \text{for } i = j \\ \dfrac{Z_i Z_j}{\| R_i - R_j \|} & \text{for } i \neq j \end{cases} \qquad (1)$$

$Z_i$ is the atomic number of atom i, and $R_i$ is its position in atomic units [7]. CM is symmetric and has as many rows and columns as there are atoms in the molecule. As we mentioned earlier, the Coulomb matrix is invariant to rotation, translation but not to permutation of its atoms. One remedy that we used in this study, is to sort these matrices by descending order with respect to the norm-2 of their columns and simultaneously permuting rows and columns accordingly. After the ordering step and given the symmetry of these matrices, it is customary to only consider their lower triangular part [6, 7], and to unfold them row-wise in a 1-dimensional (1D) vector of length L = $\sum_{i=0}^{I}(I - i)$, where I corresponds to the number of atoms of the largest molecule. In this study, the 1D vector is called the CM signal x(m,:) = $x_m[l]$, with l = 1 to L and m corresponds to molecule $\Omega_m$. For a set of M molecules, their 1D CM signals can be organized in an M×L matrix x:

$$x = \begin{bmatrix} x_{11} & x_{12} & \ldots & x_{1l} & \ldots & x_{1L} \\ x_{21} & x_{22} & \ldots & x_{2l} & \ldots & x_{2L} \\ \vdots & \vdots & \ldots & \vdots & \ldots & \vdots \\ x_{m1} & x_{m2} & \ldots & x_{ml} & \ldots & x_{mL} \\ \vdots & \ldots & \ldots & \vdots & \ldots & \vdots \\ x_{M1} & x_{M2} & \ldots & x_{Ml} & \ldots & x_{ML} \end{bmatrix} \quad (2)$$

The $m^{th}$ row of *x* represents the 1D CM signal of the $m^{th}$ molecule. Given that molecules have different number of atoms, the short ones are padded with zeros so that all the 1D CM signals have the same length L.

### 3.3    Input of the QM/ML model

Let's define X as the input to the neural network defined below. In order to test the usefulness of the AC in the prediction of the electronic properties of molecules, we have considered three different inputs and compared them against each other. The three inputs are: X = α (only the AC is used), X = x (only the CM is used), and finally X = [α x] (AC and CM are combined and used as inputs). By combining AC and CM, taking the Z-scores of X prior to its utilization as input to the ML model becomes an obvious choice. The Z-score of X will return a matrix of same size Z, where each column of Z has mean 0 and a standard deviation of 1 [18].

### 3.4    Output – Atomization Energy of Molecules

The output to the QM/ML is the atomization energy E. It quantifies the potential energy stored in all chemical bonds. As such, it is defined as the difference between the potential energy of a molecule and the sum of potential energies of its composing isolated atoms. The potential energy of a molecule is the solution to the electronic Schrödinger equation $H\Phi = E\Phi$, where H is the Hamiltonian of the molecule and $\Phi$ is the state of the system. The atomization energy of molecules are organized in an M×1 column vector $y = [y_1\ y_2\ \ldots\ y_m\ \ldots\ y_M]^T$. The superscript T indicates the transpose operator. The entry $y_m$ is a real number that corresponds to the atomization energy of the $m^{th}$ molecule.

### 3.5    Bayesian Regularized Neural Networks

Neural networks (NN) are universal function approximators that can be applied to a wide range of problems such as classification and model building. It is already a mature field within machine learning and there are many different NN paradigms. Multilayer feed-forward networks are the most popular and a large number of training algorithms have been proposed. Compared to other non-linear techniques, in multilayer NNs, the measure of similarity is learned essentially from data and implicitly given by the mapping onto increasingly many layers. In general, NNs are more flexible and make fewer assumptions about the data. However, it comes at the cost of being more difficult to train and regularize [3]. In this paper, we used the Bayesian regularization method to train our NNs [9-12].

Bayesian methods are optimal methods for solving learning problems. Any other method not approximating them should not perform as well on average. They are very useful for comparison of data models as they automatically and quantitatively embody "Occam's Razor" [19]. Complex models are automatically self-penalizing under Bayes' Rule. Bayesian methods are complementary to NNs as they overcome the tendency of an over flexible network to discover nonexistent, or overly complex, data models.

Unlike a standard back-propagation NN training method where a single set of parameters (weights, biases, etc.) are used, the Bayesian approach to NN modeling considers all possible values of network parameters weighted by the probability of each set of weights. Bayesian inference is used to determine the posterior probability distribution of weights and related properties from a prior probability distribution according to updates provided by the training set $D$ using the Bayesian regularized NN model, $H_i$. Where orthodox statistics provide several models with several different criteria for deciding which model is best, Bayesian statistics only offer one answer to a well-posed problem.

$$P(w | D, H_i) = \frac{P(D | w, H_i) P(w | H_i)}{P(D | H_i)} \tag{8}$$

Bayesian methods can simultaneously optimize the regularization constants in NNs, a process which is very laborious using cross-validation [9].

## 4　Results and Discussions

As we mentioned earlier, the version of the QM7 dataset used in this study is the one published in [7] and it is composed of M = 7102 molecules and contains up to five types of atoms: Carbon (C), Hydrogen (H), Oxygen (O), Azote (N), and Sulfur (S). Therefore the set of unique atoms is A = {C, H, N, O, S}. The matrix α is of size M×K = 7102×5. The largest molecule is made of I = 23 atoms. Thus each CM is of size 23×23 and the x matrix is of size M×L = 7102×276, because L = $\sum_{i=0}^{23}(23 - i)$ = 276. The column vector of atomization energy y is of size 7102×1. The QM7 dataset is randomly divided into 80% training and 20% testing sets. Performance is measured using the root mean square error (RMSE), **Equation 9**, the mean absolute error (MAE), **Equation 10**, and the Pearson correlation coefficient $r_{ppe}$, **Equation 11**.

$$RMSE = \sqrt{\frac{1}{M} \sum_{m=1}^{M} (y_m - y_m^e)^2} \tag{9}$$

$$MAE = \frac{1}{M} \sum_{m=1}^{M} |y_m - y_m^e| \tag{10}$$

$$r_{PP^e} = \frac{\sum_{m=1}^{M}(y_m - \bar{y})(y_m^e - \bar{y}^e)}{\sqrt{\sum_{m=1}^{M}(y_m - \bar{y})^2}\sqrt{\sum_{m=1}^{M}(y_m^e - \bar{y}^e)^2}} \quad (11)$$

### 4.1 Results

We used the Matlab implementation of the regularized Bayesian network to model the relationship between the inputs (X) and the output (y). **Figure 1** for example shows the Matlab architecture of one of the networks used.

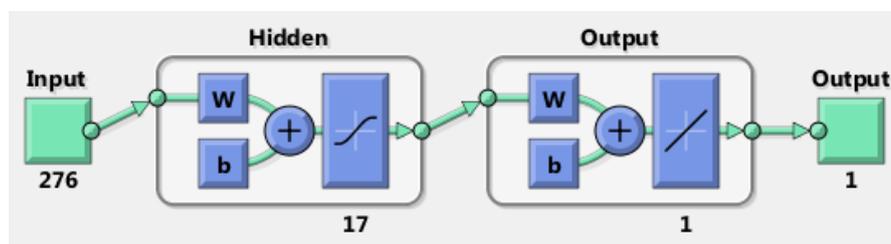

**Fig. 1.** Matlab representation of one hidden layer neural networks with 17 neurons when the CM is used as input (276 inputs and 1 output).

In the literature, there is no clear and rational approach on how to select the number of neurons and the number of hidden layers of a NN. The middle ground is usually to select an architecture that will neither under-fit nor over-fit the model. In this study we tested several architecture based on some empirical observations also coming from the literature with the goal for avoiding under-fitting and overfitting of the model. **Table 1** shows the results obtained using different NN architectures with the data partitioned into 90% training and 10% for validation and testing.

**Table 1.** Statistics of the results using four different network architectures, each trained using Bayesian regularization. MAE and RMSE are in kcal/mol.

| Network architecture | Statistics | Input | | |
|---|---|---|---|---|
| | | AC | CM | [AC CM] |
| [17] | MAE | 13.82 | 5.02 | **3.70** |
| | RMSE | 18.05 | 6.72 | **5.0** |
| | $r_{ppe}$ | 0.9967 | 0.9995 | **0.9997** |
| [16 × 8 × 4] | MAE | 13.82 | 4.83 | **3.42** |
| | RMSE | 18.05 | 6.58 | **4.73** |
| | $r_{ppe}$ | 0.9967 | 0.9996 | **0.9998** |
| [18 × 9 × 3] | MAE | 13.80 | 4.40 | **3.0** |
| | RMSE | 18.04 | 5.95 | **4.22** |
| | $r_{ppe}$ | 0.9967 | 0.9996 | **0.9998** |

The results obtained show that the association of the AC with CM significantly improved the prediction accuracy. For example, with the three hidden layer network [18 × 9 × 3], the MAE goes from 13.80 kcal/mol when only the AC is used, to 4.40 kcal/mol when only the CM is used down to 3.0 kcal/mol when the AC is combined with CM. These results suggest that AC represents an interesting feature for the predictions of the electronic properties of molecules. Furthermore, the MAE = 3.0 kcal/mol obtained is lower than the MAE of 9.9 kcal/mol [6, 7] using kernel ridge regression and MAE of 3.51 kcal/mol obtained in [1, 3] using a multilayer NN associated with random coulomb matrices and a binarization scheme to augment the data in order to use a more complex multilayer neural network than the one used in this study. Clearly, the QM7 dataset result of [1, 3] is improved by a factor difference of 0.5kcal/mol in this study. Our result is close to the acceptable 1 kcal/mol chemical accuracy.

### 4.2  Discussions

Predicting molecular energies quickly and accurately across the CCS is an important problem as the QM calculations take days and do not scale well to more complex molecules. ML is a good candidate for solving this problem as it encourages the framework to focus on solving the problem of interest rather than solving the more general Schrödinger equations. In this paper, we have developed further the learning-from-scratch approach initiated [6] and provided a new ingredients for learning a successful mapping between raw molecular geometries and atomization energies. Our results suggest important discoveries and open new venues for future research.

**Atomicity, atom counts or atomic composition represents an interesting feature for QM/ML models**.  Atomic composition (AC), i.e. atom counts of each type in a molecule is a representation that does not contain any molecular structural information. But our analysis suggests a correlation between the AC representation and the atomization energy. The combination of AC and CM yield a new molecular representation which inherits all the properties of the CM representation. Even though the AC representation is not unique (case of isomers as we mentioned earlier), by combining it with the CM, the pair [AC CM] inherit all the properties of CM and becomes a representation that is uniquely defined, invariant to rotation, translation and re-indexing of the atoms, given that the CM had already been sorted in decreasing order to tackle the non-invariance to atom re-indexing.

**Bayesian regularized neural networks are suited for molecular data**.  The Bayesian regularization approach used in this study seems to fit molecular data very well. Similar observation was made in [13] when developing quantitative structure activity relationship (QSAR) model of compounds active at the benzodiazepine and muscarinic receptors. The results obtained here further prove the point that Bayesian regularized neural networks possess several properties useful for the analysis of molecular data. One advantage of the Bayesian regularized neural networks is that the number of effective parameters used in the model is less than the number of weights, as some weights do not contribute to the models. This minimizes the likelihood of overfitting.

The concerns about overfitting and overtraining are also removed by this method so that the production of a definitive and reproducible model is attained [9-12].

## 5      Conclusions

In this study, we show that by combining the atomic composition of molecules with their Coulomb matrix representation, the output of the quantum mechanics machine learning model can be significantly improved. Using the QM7 dataset as a test case, our results show a decrease by a difference of 1.5 kcal/mol when Coulomb matrix is combined with the atomic composition compared to when Coulomb matrix is used alone, and also by a difference of 0.5 Kcal/mol compared to well-known results from the literature. These results suggest that the atomic composition of molecules contain interesting information useful for quantum mechanics machine learning model and should not be neglected.